\documentclass[namedreferences]{SolarPhysics}
\usepackage[optionalrh,solaenum,natbib]{spr-sola-addons} 
\usepackage{graphicx}                    
\usepackage{color}                       
\usepackage{url}                         

\begin{document}

\begin{article}

\begin{opening}

\title{The Multiple Continuum Components in the White Light Flare of 16 January 2009
 on the dM4.5e Star YZ CMi }

%
\author{A.F.~\surname{Kowalski}$^{1}$\sep
             S.L.~\surname{Hawley}$^{1}$\sep
              J.A.~\surname{Holtzman}$^{2}$\sep    
              J.P.~\surname{Wisniewski}$^{1,3}$\sep 
              E.J.~\surname{Hilton}$^{1}$      
       }

%
\runningauthor{A.F.~Kowalski \emph{et al.}}
\runningtitle{The Multiple White Light Continuum Components}

%
  \institute{$^{1}$ Astronomy Department, University of Washington,
    Box 351580, Seattle, WA 98195, USA. email: \url{adamfk@u.washington.edu}\\ 
      $^{2}$ Department of Astronomy, New Mexico
               State University, Box 30001, Las Cruces, NM 88003, USA \\
      $^{3}$ NSF Astronomy \& Astrophysics Postdoctoral Fellow. \\
             }

\begin{abstract}

The white light during M dwarf flares has long been known 
to exhibit the broadband shape of a $T\approx10\,000$ K blackbody, and the 
white light in solar flares is thought to arise primarily from Hydrogen
recombination.  Yet, a current lack of broad wavelength coverage
solar-flare spectra in the optical/near-UV 
prohibits a direct comparison of the continuum properties to
determine if they are indeed so different.  New spectroscopic observations of 
a secondary flare during the decay of a megaflare on the dM4.5e star YZ CMi 
have revealed multiple components in the white-light continuum of stellar 
flares, including both a blackbody-like spectrum and a hydrogen recombination 
spectrum.  One of the most surprising findings is that these two
components are anti-correlated in their temporal evolution.  We combine initial 
phenomenological modeling of the continuum components with spectra from 
radiative-hydrodynamic models to show that continuum veiling gives rise to 
the measured anti-correlation.  This modeling allows us to use the components' 
inferred properties to predict how a similar spatially resolved, 
multiple-component white-light continuum might appear using analogies
to several solar flare phenomena.  
We also compare the properties of the optical stellar flare white
light to Ellerman bombs on the Sun.

\end{abstract}

%

\end{opening}

%

\section{Introduction}
     \label{intro} 
In both solar and stellar flares, the near-UV and optical (white-light) continuum emission is an 
energetically important but unexplained phenomenon.  On the Sun, the white-light
continuum appears in small regions of transient emission that are spatially and temporally
coincident with hard X-ray bursts \citep{Rust1975,Hudson1992,NeidigKane1993,Fletcher2007}.  This relation suggests that the origin of
the white light is related to the energy deposited in the lower atmosphere by nonthermal electrons
accelerated during flares.  Broad wavelength coverage spectral
observations are sparse and date back to several large solar flares from the
1970s and 1980s \citep{Machado1974, Hiei1982, Neidig1983,
  DonatiFalchi1984}.  These
spectra are consistent with continua
arising primarily from the hydrogen Balmer continuum and H$^{-}$ emission.
  
Whereas the largest solar flares emit $< 10^{32}$ ergs in the white-light
continuum and last not much longer than ten minutes
\citep[e.g.,][]{Neidig1994}, the white-light emission on active lower-mass M dwarfs can reach
$> 10^{34}$ ergs and persist for hours \citep[hereafter K10a]{HawleyPettersen1991,
Kowalski2010a}.  More is
known about the spectral shape of the white light during M dwarf
flares (many spectrographs can easily obtain
low-resolution, broad wavelength spectra of stellar flares, but it is difficult to place a spectrograph slit over solar
white-light kernels, which are intermittent and largely unpredictable), which have been studied using broadband
colors \citep{Hawley2003, Zhilyaev2007} and optical/NUV
spectra \citep[K10a]{HawleyPettersen1991, Eason1992, GarciaAlvarez2002,
  Fuhrmeister2008}.  
In contrast to solar observations, the spectral
shape during M dwarf flares suggests a hot blackbody with
temperatures
between $\approx8500 - 11\,000$ K.  Areal coverages of this component are
typically $ < 0.1$\% of the visible stellar hemisphere, which implies
a compact geometry like those observed in white light at the 
footpoints of flare arcades on the Sun.  Although this
blackbody component seems to be nearly ubiquitous during (large)
stellar flares, it is not predicted even by the most recent 1D radiative
hydrodynamic (RHD) flare models \citep{Allred2006} produced with the
RADYN code \citep{Carlsson1994,Carlsson1995,Carlsson1997}.

The ``megaflare'' of UT 16 January 2009 on the dM4.5e star YZ CMi is one of the longest
lasting and most energetic flares observed on a low-mass single star.
Low-resolution spectra (3350\,--\,9260\AA) 
were obtained in the
flare's decay phase, which was elevated between 15 and 37 times the quiescent
level and contained many secondary peaks.  More than 160 spectra were
obtained over 1.3 hours, and simultaneous U-band photometric observations of the entire
flare event were provided by the NMSU 1m telescope.  A detailed
description of the observations and data reduction is given in K10a.

In K10a, two continuum components
were necessary to fit the blue
(3350\,--\,5500\AA) spectra: a hydrogen Balmer
continuum (BaC) component as predicted by the RHD models of 
\cite{Allred2006} and a $T\approx10\,000$ K blackbody component.
An intriguing anti-correlation was found between the
temporal evolution
of these two
components: the blackbody emission increased when the BaC decreased,
and \emph{vice versa} (see Figure 1d of K10a).  In this article, we revisit
this anti-correlation and provide an explanation for it using the
phenomenological models of the secondary flare spectra from
\citet[hereafter K10b]{Kowalski2010b}.  Finally, we show how each component of this flare
might appear in the context of a solar flare ``arcade''.

\section{Anti-Correlated Continuum Components and Continuum Veiling}
     \label{anticor} 
\begin{figure}[!ht]
 \centerline{\includegraphics[width=0.75\textwidth,angle=90]{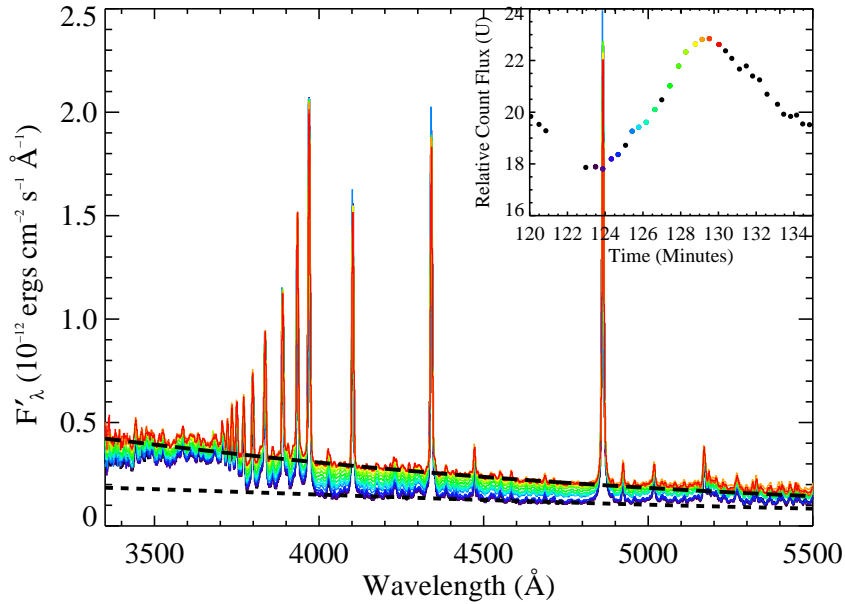}}
 \caption{A series of 16 flare spectra obtained prior to the onset and through
   the peak of a secondary flare.  The quiescent spectrum from 24
   November 2008 has been subtracted, as in K10a.  The U-band light curve (inset) is
   color-coded to the spectrum obtained closest in time.   The
   best-fit blackbody curves to the black- and red-colored spectra
   are shown as the short-dashed ($T\approx 10\,400$ K) and long-dashed lines
   ($T\approx 13\,000$ K), respectively.
 }
\label{F-riseseq}
 \end{figure}

The anti-correlation between the blackbody and BaC components (K10a) can be understood qualitatively using Figure
\ref{F-riseseq}, where we show the spectral
evolution of the total ``flare-only'' flux (denoted here as $F_{\lambda}^{\prime}$) during the rise phase of the secondary flare at $t \approx 130$
minutes.  The spectra
are color-coded to the nearest (in time) U-band measurement in the inset
panel.  At times prior to and near the beginning of the secondary
flare (black, purple, and dark blue spectra), two distinct continuum
components are clearly present in the spectra.  The best-fit blackbody
(short-dashed line) accounts for most 
continuum emission at $\lambda > 4000$\AA, whereas the BaC emission above 
the blackbody is conspicuous at $\lambda < 3750$\AA.  During the rise 
and at the peak of the secondary flare (green, yellow, and red
spectra), the BaC component seemingly disappears
and the best-fit blackbody (long-dashed line) can fit the continuum
shape throughout
 the entire wavelength range.  K10a 
showed that the H$\gamma$\ line flux exhibits an anti-correlated
relation with the blackbody component.  This effect is also present in Figure
\ref{F-riseseq}:   In the red (secondary flare peak) spectrum, the continuum at $\lambda \approx
4200$\AA\ is highest, yet the peaks of the hydrogen Balmer lines are
lowest.  

The secondary flares at $t \approx 95$\ minutes and $t \approx 130$\ minutes are
events during which the blackbody flux becomes stronger while the
BaC flux becomes weaker.  K10a quantified this as an increasing
filling factor
(areal coverage; percent of stellar disk) of the blackbody (with constant temperature,
$T = 10\,000$ K) during the rise phase of each secondary flare.  In Figure \ref{F-riseseq}, we
present an alternative interpretation.  The blackbody
curves (dashed lines) were fit to
the spectra by allowing both the temperature and filling factor to vary.
The best-fit blackbody temperatures and filling factors
are $T\approx 10\,400$ K and $X_{\mathrm{BB}} \approx 0.1$\%
(at $t = 122.9$ minutes; short-dashed line) and $T \approx 13\,000$ K and $X_{\mathrm{BB}} \approx
0.1$\% (at $t = 130$ minutes; long-dashed line).  Strikingly, if both 
parameters are allowed to vary when fitting a blackbody function to these
\emph{total} flare spectra, the temperature increases by $\approx2500$ K while the filling
factor remains approximately constant.  Fitting a blackbody to the \emph{total} flare
spectrum (either by holding $T$ constant, or by allowing $X$ and $T$ to
vary) gives 
only the \emph{average} properties of the entire flaring region at that
time.  We next show that these interpretations can be improved by isolating
the 
newly-formed flare emission.

K10b found that the isolated flare emission (denoted here as $F_{\lambda}^{\prime\prime}$) during the 
secondary flare's rise phase resembles the spectrum of a hot star, with the
defining features being the 
hydrogen Balmer continuum and lines in \emph{absorption} and a steeply
rising continuum towards the blue at $\lambda > 4000$\AA\ (Figure 1b of
K10b shows that the new flare emission, obtained by subtracting the 
pre-secondary flare spectrum (average of three black and purple spectra around $t =
123.4$ minutes) from the average of two green-colored spectra around $t =
126.5$ minutes
in Figure \ref{F-riseseq} of this article, is very similar to the
spectrum of the A0 star
Vega).  The
observed anti-correlation between the continuum components in Figure
\ref{F-riseseq} is a result of a `hot star spectrum'
forming during the secondary flare. The hot star  (``blackbody-like'') spectrum causes an increase in the continuum at $\lambda \approx
4200$\AA\ by an amount, $F_{4200}^{\prime\prime}$, whereas an increase in the continuum at $\lambda \approx 3500$\AA\
occurs by only $\approx0.6 \times F_{4200}^{\prime\prime}$.  In other words, the flux in the continuum
on both sides of the Balmer jump increases, but
the continuum at $\lambda \approx 4200$\AA\ increases by a larger
amount.   Thus, the apparent decrease in the total amount of BaC
in emission from $t = 123$ minutes to $130$ minutes occurs as a result
of `continuum veiling' (similar to the continuum
  veiling observed for
 accreting T Tauri stars -- see, \emph{e.g.},
 \cite{Hartigan1989,Hessman1997,Herczeg2008}).

\section{Combining Continuum Components using Phenomenological Hot
  Spot Models}
     \label{phenom addition} 

\begin{figure}[!ht]
 \centerline{\includegraphics[width=0.75\textwidth,angle=90]{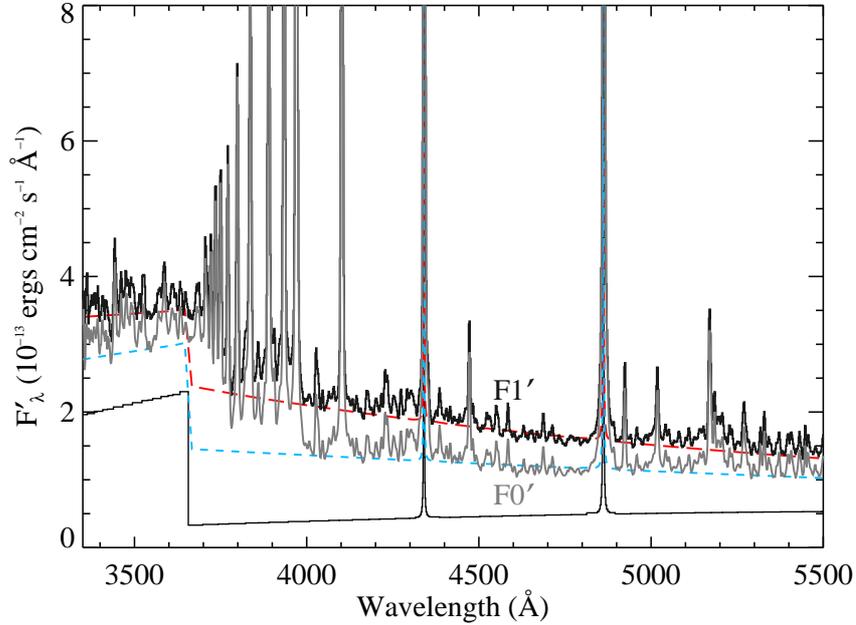}}
 \caption{Flare spectra at $t = 123.4$ minutes and $126.5$ minutes from Figure \ref{F-riseseq} are shown in grey and black, respectively.
   The composite model spectra are shown in blue, short dashes ($F_{\lambda,\mathrm{RHDF11}}
   \times X_{\mathrm{RHDF11}} + F_{\lambda,\mathrm{HS1}} \times
   X_{\mathrm{HS1}}$) and red, long dashes ($F_{\lambda,\mathrm{RHDF11}}
   \times X_{\mathrm{RHDF11}} + F_{\lambda,\mathrm{HS1}} \times X_{\mathrm{HS1}} +
   F_{\lambda,\mathrm{HS2}} \times X_{\mathrm{HS2}}$).  The
   RHDF11 spectrum is shown as the thin black line. The continuum veiling
   effect is apparent from the different heights of the Balmer
   discontinuity at $\lambda = 3646$\AA. }
\label{F-sq_add}
 \end{figure}

In K10b, the $F_{\lambda}^{\prime\prime}$ emission was modelled phenomenologically with
the static radiative transfer code, RH \citep{Uitenbroek2001}, as a
temperature bump (``hot spot'') with peak temperature $T = 20\,000$ K, placed near the
photosphere (below the temperature minimum) of the quiescent M dwarf atmosphere.  Here, we use a sum of individual hot spots and the 
RHD model spectrum (hereafter RHDF11) of \cite{Allred2006} to model
the \emph{total} flare emission ($F_{\lambda}^{\prime}$) at two times
during the megaflare on YZ CMi.  Figure \ref{F-sq_add} shows 
flare spectra from Figure \ref{F-riseseq} averaged around
$t = 123.4$ minutes ($F0^{\prime}$; grey) and at $t = 126.5$ minutes
($F1^{\prime}$; black).  These are the spectra
corresponding to times immediately before and nearly
half-way up the rise phase of the secondary flare, respectively (\emph{i.e.},
the
same two spectra presented in Figure 1a of K10b but with the quiescent level subtracted).

To model $F0^{\prime}$, we add the RHDF11 spectrum and a
hot spot (HS1) with $T_{\mathrm{max}} = 12\,000$ K (keeping the other parameters
the same as described in
K10b) with a ratio of filling factors of
$10:1$ and $X_{\mathrm{RHDF11}} = 1.2$\%, as found in K10a.  The total model spectrum is shown as the
light blue (short dashes) curve in Figure \ref{F-sq_add}.   As in K10b, we model the
secondary flare as a hot spot
(HS2) with $T_{\mathrm{max}} = 20\,000$ K.  Adding HS2 to RHDF11 and HS1 gives
the red (long dashes) spectrum in Figure \ref{F-sq_add}.  The areal coverage of HS2 is
0.46 as large as the areal coverage of HS1.  These model spectra
match
the observed continuum levels at all wavelengths in the figure.  Moreover, the
continuum veiling / anti-correlation effect is readily apparent:  the height of the
Balmer jump at $\lambda = 3646$\AA\ decreases from black (no hot spots)
to blue (one hot spot) to red (two hot spots).  The model fluxes are
slightly lower than the observed fluxes at $\lambda < 3750$\AA.  
At these wavelengths, there is a forest of metallic lines (\emph{e.g.},
Fe \textsc{i}, Fe \textsc{ii}) that are blended in our low-resolution spectra; our model is satisfactory in matching the
\emph{underlying} level which is likely closer to the actual
level of the BaC.  Note that in K10a, we intentionally used only the BaC from the
RHDF11 spectrum to model the continuum, whereas the RHDF11
predictions for the Paschen continuum and photospheric-backwarming components are
included in the total fluxes in this work.

The origin of the hot spots is unknown, as they are not
predicted by self-consistent radiative hydrodynamic models that employ a solar-type
non-thermal electron heating function \citep{Allred2006}.  
In the proposed continuum model, we use the fewest number of components
necessary to fit the overall shape and reproduce the
anti-correlation.  However, HS1 may represent
a sum of individual decaying hot spots from previously heated flare
regions (see below).  We are working to produce a grid of phenomenological models which will be
used to constrain the column mass of the hot spots, the
detailed temperature evolution, and the uniqueness of the continuum
fit.  Radiative hydrodynamic models (with RADYN) of the gradual phase are also
forthcoming and will provide a more accurate representation of the
BaC and photospheric backwarming in the decay phase.

\section{The Solar Analogy}
     \label{cartoonsection} 

\begin{figure}[!ht]
 \centerline{\includegraphics[width=0.75\textwidth]{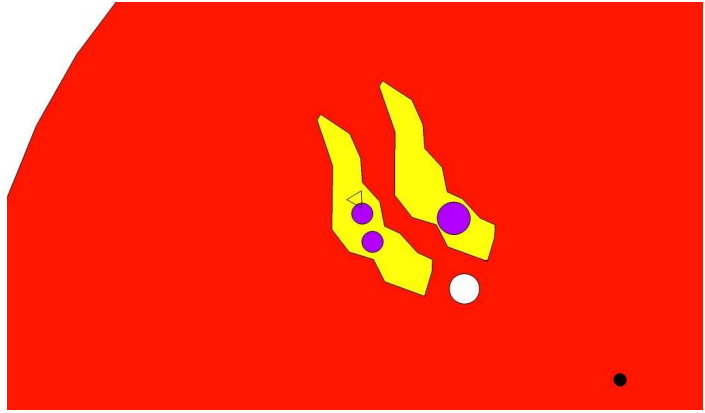}}
 \caption{Graphic with continuum components (areas from $t \approx
   126.5$ minutes to scale) as they might
   appear in a spatially resolved observation.  The BaC emitting
   region (yellow) resembles a two-ribbon structure (shown here as
   symmetric for simplicity) in a thin
   layer of the heated mid-to-upper chromosphere of previously reconnected
   magnetic loops.  HS1 is
   shown as a collection of previously formed hot spots (purple) and
   HS2 is the proposed newly-formed continuum emitting region (white).
   The triangle indicates the assumed location of the inital flare peak,
   which generated a disturbance in the lower atmosphere that propagated into
   the surrounding active region and triggered the hot spots.  The black circle
   helps orient the reader to the center of the star, which has a
   radius $0.3 R_{\odot}$. The flare region is
   placed at an arbitrary location on the surface.  Several aspects of
   this cartoon were inspired by observations of solar flare arcades,
   such as \cite{Fletcher2001}.}
\label{F-cartoon}
 \end{figure}

Figure \ref{F-cartoon} shows how a spatially resolved observation
of the YZ CMi megaflare might have appeared.  We use the 
continuum
 components and filling factors to make analogies to several 
flare structures and phenomena observed in large solar flare
arcades.  The complex morphology of the U-band light curve leads
us to speculate that the YZ CMi megaflare involved a large arcade, or
 several large arcades of flare loops.  The main features of Figure
 \ref{F-cartoon}
 are the following:

\begin{itemize}
\item \textbf{BaC (yellow ribbons)}:  Throughout the spectral observations
  ($72$ minutes $ < t < 149$ minutes), the hydrogen BaC and lines were highly elevated
 and decreasing (likely from the initial flare peaks in the U-band light
 curve), implying that this
 emission had originated from the footpoints of a previously heated 
magnetic arcade in the flaring
chromosphere.  These may manifest as a complex of flare ribbons, as is
commonly observed in H$\alpha$ during solar flares
\citep[\emph{e.g.},][]{Rust1975, Berlicki2004, Bala2010}.  In some solar
flares \citep[\emph{e.g.},][]{Neidig1983}, the BaC appears to have a spatial morphology
that is more compact than an extended H$\alpha$ ribbon.
Spectroscopic observations are needed to compare the plasma properties and
conditions of BaC and Balmer line emitting ribbons
and kernels.

\item \textbf{HS1 (purple spots)}:  Immediately prior to the secondary flare beginning at $t \approx
  123$ minutes, 
  a series of phenomenological hot
  spots (HS1) are present near the photosphere.  These hot spots were
  formed during the previous secondary flares (\emph{i.e.}, at $t \approx 65$
  minutes, $\approx 95$ minutes; see K10a), and they are emitting from a total source
  size that is $\approx1/10$ as
  large as the area of the
  chromospheric flare region (\emph{e.g.}, H$\alpha$ ribbons).  The spectra
  of these hot spots have the hydrogen BaC and lines in
  absorption.   They might be similar to
  the compact white-light kernels during solar flares, as in
  \cite{Wang2007}, \cite{Fletcher2007}, \cite{Jess2008}, or they may be
  similar to Ellerman-bomb phenomena (see Section \ref{ellerman}).  Also emitting from the photosphere is a
  larger region heated from chromospheric (BaC) backwarming; we assume the size of
  this backwarmed region is the same size as the flaring chromosphere.  

\item \textbf{HS2 (white spot)}:  The secondary flare at $t \approx 130$ minutes is the result of the formation
  of a new hot spot (HS2), hotter and smaller than HS1 but at the same
  column mass.  
  At this time, we see a
  sudden decline in the BaC flux.  When all of these components are
  unresolved, as in our stellar spectra, continuum veiling
  gives rise to the observed anti-correlation.  We have
  placed HS2 assuming it was triggered by a disturbance induced by the huge
  initial flare peak at $t \approx 28$ minutes.  The time-evolution of the
  H$\gamma$ and BaC fluxes in Figure 1d of K10a indicates an apparent
  lack of 
  new hydrogen Balmer line emitting-regions during the secondary
  U-band peaks (we cannot definitively determine
whether there is a newly-formed hydrogen Balmer-emitting
(chromospheric ribbon) component cospatial
with the hotspot because the observations are unresolved).
 Therefore, the 
  disturbance likely propagated through the lower atmosphere, below the
  height of hydrogen Balmer line formation (upper chromosphere;
  J. Allred, private communication 2010). 
  Using a range of sound speeds
  in the lower atmosphere for the
  speed of the disturbance ($\approx 5-10$
  km s$^{-1}$), we find that HS2 is located at a distance that is
  approximately $30-60$ Mm (R$_{\mbox{YZ CMi}} \approx 200$ Mm) from the site of
  the initial flare event.\footnote{The white-light and hard X-ray footpoints have been observed to propagate along the
  polarity inversion line during large solar flares such as the famous
  14 July 2000 flare (\cite{Fletcher2001, Kosovichev2001, Qiu2010}, but the spatial location
  of these kernels appears to change much faster, $\approx 170-200$ km
  s$^{-1}$.}  
\end{itemize}

The composite graphical model is preliminary (see Section \ref{conclusions}) and requires comparison to other
complex flare events on dMe stars but especially to solar flares where we
can spatially resolve each continuum component.  Our group is currently working
to obtain solar flare data that can be used to test the YZ CMi flare
model using DST/ROSA and employing custom continuum filters \citep{Jess2010,Kowalski2011}.  

\subsection{Are the Secondary Flares Stellar Ellerman Bombs?}
\label{ellerman}
Ellerman bombs are transient, compact phenomena observed near
evolving or emerging magnetic fields in solar active
regions \citep[and references therein]{Ellerman1917, Severny1968, Georgoulis2002}.  A typical signature of
Ellerman bombs is emission in the wings 
and absorption in the core of H$\alpha$ relative to the nearby 
plage intensity \citep[\emph{e.g.},][]{Koval1970, Bruzek1972, Fang2006}.  The time-evolution
properties include 
mean lifetimes of $\approx10-20$ minutes and fine-structure variations
\citep{Kurokawa1982, Qiu2000}.  In contrast to typical
white-light flares, Ellerman bombs have symmetric light curves with
similar rise and decay times \citep{Payne1993, Qiu2000, Jess2010b}.
The Ellerman-bomb mechanism is not fully understood but has been
attributed
to magnetic reconnection in the low chromosphere \citep[\emph{e.g.},][]{Georgoulis2002}.
 
The secondary flares during the YZ CMi megaflare
exhibit several similarities to Ellerman-bomb phenomena on the Sun.  Ellerman bombs have also
been modeled phenomenologically as temperature
bumps at or below the solar temperature minimum region
\citep{Fang2006, Berlicki2010}.  
The secondary YZ CMi flares have longer rise times ($\approx$ two\,--\,five minutes)
and are much more
symmetric about the peak compared 
to other white
light flares with similar total energy on YZ CMi \citep[$\Delta
t_{\mathrm{rise}}\approx 0.5 - 1.8$ minutes; ][]{Moffett1974,vandenoord1996}.
The absorption features of the blackbody-like continuum
component are similar to the line-center absorption observed in
H$\alpha$ and Ca \textsc{ii} during Ellerman bombs; unfortunately, our observations do not have
sufficient spectral resolution to separate line-center and wing profiles.
The preliminary finding (Section \ref{cartoonsection}) that the blackbody-like continuum
component does not contain hydrogen-line emission may be consistent with magnetic reconnection
taking place in the low atmosphere.  However, in contrast to
solar Ellerman bombs, which have been observed as a microflare trigger
\citep{Jess2010b}, the secondary flares are possibly a
\emph{consequence} of the enormous
YZ CMi flare peak event.

\section{Summary and Future Work}
     \label{conclusions} 
The time-resolved continuum data obtained during a
megaflare on the dM4.5e star YZ CMi demonstrate the power of
broad-wavelength
coverage, low-resolution spectra, which are unfortunately not
currently available in the optical/near-UV for solar flares.  In this manuscript, we show that the blackbody-like
component (hot star-like emission) of the white-light continuum dominates the spectra during
the secondary flares while the BaC (and Balmer lines) become less
important; 
 this observed anti-correlation 
is explained as continuum veiling.  
We combine the phenomenological models of the blackbody-like component from
K10b and the \cite{Allred2006} RHD model spectrum of the BaC to
reproduce the total flare emission at two times during the flare.  The
filling factors for the decaying BaC($+$backwarming) component, a previously
heated hot-spot component, and a newly heated hot spot component are in the ratio of
$\approx10:1:0.5$.  These areal coverages allow a comparison of each
component to be made with observed solar-flare structures in large eruptive
flare arcades.  Although we generally assume that stellar-flare phenomena are
simply ``scaled-up'' versions of solar-flare
phenomena, one should not exclude the possibility that stellar flares 
might have fundamental differences in the white-light
continuum, as the energies and timescales of dMe flares can be
orders
of magnitude larger than solar flares.  New solar-flare observations
are needed to test the existence of the
blackbody-like component and to better understand the properties of the
BaC, which could be fully ``unveiled'' using spatially resolved solar observations. 

We provide evidence that the blackbody-like component has several similar properties to solar Ellerman
bombs.  A few solar flares have been known to exhibit the spectral
features \citep{Svestka1963}
 and velocity characteristics \citep{Kosovichev2001} of Ellerman bombs.
Additional intensity-calibrated continuum measurements of Ellerman bombs and
white-light flare kernels on the Sun, such as with DST/ROSA, would
help illuminate the differences between these events and provide a comparison to
stellar spectra of typical white-light flares and megaflare-size events.

Several aspects of the phenomenological models presented in this work
are being improved.  In addition to modeling hydrogen with more
levels and including metallic transitions and molecular species, a
more accurate consideration of charge balance is underway.  The
correct treatment of charge balance in a modified atmosphere is
complicated by non-LTE ionization, but a new version of RH has been
provided by H. Uitenbroek to account for this; the authors are
currently working on a new suite of hot-spot models.  

%

%
 \begin{acks}
This work was first presented at `The Origin, Evolution,
and Diagnosis of Solar Flare Magnetic Fields and Plasmas:  Honoring 
the Contributions of Dick Canfield', a conference that took place from
9\,--\,11 August 2010 at HAO/NCAR in Boulder, CO.  AFK thanks the
organizers
of this conference for generous travel assistance and acknowledges
support from NSF grant AST 0807205.
We gratefully thank H. Uitenbroek for many useful discussions and for our
use of the RH code; P. Heinzel, M. Varady, and D. Jess for illuminating
conversations about
Ellerman bombs; and J. Allred for allowing us to use the detailed
output of his RADYN flare models.  
We also acknowledge Google for our use of its SketchUp application.  
Based on observations obtained with the Apache Point Observatory 3.5 m
Telescope, which is owned and operated by the Astrophysical Research
Consortium.  
 \end{acks}

%
\bibliographystyle{spr-mp-sola-cnd} 
 \bibliography{KowalskiSoPh.bib}  
%

\end{article} 
\end{document}